\documentclass [a4paper]{article}
\usepackage{epsfig}
\usepackage{ulem}

\usepackage{color}

\usepackage{epstopdf}
\usepackage{epstopdf}

\begin{document}
\noindent
{\bf Anu Venugopalan is on the faculty of the School of Basic and Applied Sciences, GGS Indraprastha University, Delhi. Her primary research interests are in the areas of Foundations of Quantum mechanics, Quantum Optics and Quantum Information.}

\bigskip
\noindent
{\bf Address for Correspondence: \\

\noindent
University School of Basic and Applied Sciences,\\ 
GGS Indraprastha University\\
Kashmere Gate, Delhi - 11 0 453\\
\bigskip
\noindent
e-mail: anu.venugopalan@gmail.com \\ 

\pagebreak
\title{Quantum Interference of Molecules - Probing the Wave Nature of Matter}
\author{Anu Venugopalan}
\maketitle

\begin{abstract}
The double slit interference experiment has been famously described by Richard Feynman as containing the "only mystery of quantum mechanics". The history of quantum mechanics is intimately linked with the discovery of the dual nature of matter and radiation. While the double slit experiment for light is easily undertsood in terms of its wave nature, the very same experiment for particles like the electron is somewhat more difficult to comprehend. By the 1920s it was firmly established that electrons have a wave nature. However, for a very long time, most discussions pertaining to interference experiments for particles were merely gedanken experiments. It took almost six decades after the establishment of its wave nature to carry out a 'double slit interference' experiment for electrons. This set the stage for interference experiments with larger particles. In the last decade there has been spectacular progress in matter-wave interefernce experiments. Today, molecules with over a hundred atoms can be made to interfere. In the following we discuss some of these exciting developments which probe new regimes of Nature, bringing us closer to the heart of quantum mechanics and its hidden mysteries.
\end{abstract}

\begin{flushleft}
{\bf Keywords} 
\end{flushleft}
{\bf matter waves, wave-particle duality, quantum interference, de Broglie wavelength, electron interference, fringes, quantum-classical boundary, decoherence}

\section{Introduction: The dual nature of radiation and matter and the birth of quantum mechanics}
At the turn of the last century, there were several experimental observations which could not be explained in terms of the of the established laws of classical physics and called for a radically different way of thinking. This led to the development of quantum mechanics, which is today regarded as the fundamental theory of Nature and the most elegant tool for describing the physics of the micro world. Some key events and developments that set the stage for the coming of quantum mechanics were associated with the black body radiation spectrum (Planck, 1901), the photoelectric effect (Einstein, 1905), the model of the atom (Rutherford,  1911), atomic spectra (Bohr, 1913), scattering of photons off electrons (Compton, 1922), the exclusion principle (Pauli, 1922), the hypothesis of matter waves (de Broglie 1925) and the experimental confirmation of the existence of matter waves (Davisson and Germer 1927).

The birth of quantum mechanics is intimately linked with discoveries relating to the nature of light. Theories relating to the nature of light have a long and chequered history. Is  light a wave or is it made up of particles? The earliest theory on the nature of light goes back to the  corpuscular theory of Newton in 1704. Though Christian Huygens had proposed the wave theory of light in 1690, Newton's corpuscular theory, according to which light is composed of tiny particles or corpuscles, was the favoured one for over a hundred years - a consequence of Newton's  towering presence and authority in the scientific community at that time. In 1801, Thomas Young performed an experiment with light where a beam of light was passed through two parallel slits in an opaque screen and formed a pattern of alternating light and dark bands on a screen beyond - this we know as interference - a phenomenon which is associated with waves. Later, other important experiments on diffraction and interference of light were also done, notably by Fresnel (1814) and others that could only be interpreted in terms of the wave theory for light. In the face of such irrefutable experimental evidence, the wave theory became the dominant and  accepted theory of the nature of light in the 19th century. In 1864 James Clerk Maxwell showed that electric and magnetic fields propagated together and that the speed of these electromagnetic waves was identical to the speed of light. It became clear at that point that light is a form of electromagnetic radiation. Maxwell's theory was confirmed experimentally with the discovery of radio waves by Heinrich Hertz in 1886. An experiment performed by Taylor in 1909 showed that even the weakest light source - equivalent to "a candle burning at a distance slightly exceeding a mile" - could lead to interference fringes. This led to Dirac's famous statement that "each photon then interferes only with itself". However, the wave nature of light was not the final word in this debate - there was experimental evidence- the photoelectric effect- which clearly needed an alternative interpretation. The discovery of the photoelectric effect and its explanation by Einstein firmly established that light (radiation) has a dual nature. In 1924, Louis de Broglie put forth the hypothesis that matter has a wave nature and the now famous de Broglie relation connects the wavelength, $\lambda$, of a particle with its momentum, $p$:
\begin{equation}
\lambda=\frac{h}{p},
\end{equation}
$h$ being Planck's constant. While this wavelength would be extremely small for large objects, particles like electrons have a wavelength which could be large enough to give observable effects. In 1927 Clinton Davisson and Lester Germer observed the diffraction of electron beams from a nickel crystal - demonstrating the wave-like properties of particles for the first time - and George (G P) Thompson did the same with thin films of celluloid and other materials shortly afterwards. Davisson and Thomson shared the 1937 Nobel prize for "discovery of the interference phenomena arising when crystals are exposed to electronic beams". Their work was a landmark result in the the development of quantum theory as it provided the critical confirmation of Louis de Broglie's hypothesis. Now that the wave nature of electrons was established, it remained to be see if they indeed showed the classic signature of the quantum world - the double slit interference effect, which would be the most satisfying confirmation of the dual nature of electrons as predicted by quantum theory. 

Most students of physics are familar with Richard Feynman's famous description of the double- slit expriment which captures the dual nature of matter as described by quantum mechanics. Feynman goes to great lengths to explain the apparently paradoxical phenomenon by using the example of 'bullets' and 'single electrons'. The most baffling conclusion of this experiment is that even when there is only one electron (or photon) fired at the double slit, there will be an interference pattern on the screen - something that can only be understood by the  quantum mechanical description in terms of wavefunctions, linear superposition and probability amplitudes. In the quantum mechanical description the wave and particle aspects are inseparable and it is as though the electron went through both slits simultaneously and the amplitudes for these combine at the screen to give us the interference pattern. Here lies the great 'mystery' of quantum mechanics, its predictions being completely in contrast to our  cherished classical 'common-sense' perceptions.

While most people have heard about Young's double-slit experiment for light - not many know about the experiments for electrons. Who actually performed the double-slit interference experiment for single electrons and when?
The earliest experiment can be attributed to Ladislaus Laszlo Marton of the US National Bureau of Standards (now NIST) in Washington, DC , who demonstrated electron interference in the early 1950s. However, his experiment was in a Mach-Zehnder rather than a double-slit geometry. A few years later Gottfried M\"{o}llenstedt and Heinrich D\"{u}ker  of the University of T\"{u}ingen in Germany used an electron biprism to split an electron beam into two components and observe interference between them. In 1961 Claus J\"{o}nsson performed an actual double-slit experiment with electrons for the first time. Finally - in 1989, the now famous experiment involving single electrons was performed by by Akira Tonomura and co-workers at Hitachi, Japan,  where they observed the build up of the fringe pattern with a very weak electron source and an electron biprism.
For details on the history of these interference experiments with electrons the interested reader is referred to an informative article in Physics World listed at the end of this article. In the following we will briefly review the Davisson and Germer experiment and discuss the classic double-slit experiment for a single electron performed by Tonomura et al. We will then discuss recent experiments which carry  interference experiments to a completely new level - molecules with as many as 100 atoms showing quantum interference! These are stunning experiments with the largest objects ever to show quantum interference - probing a hitherto inaccesible regime which lies in the twilight zone between the classical and quantum worlds. This is an an area of fundamental scientific curiosity and perhaps holds the key to a myriad of possibilities of practical importance.

\section{The Davisson and Germer experiment}

Clinton Davission and Lester Germer performed the conclusive experimental test of Louis de Broglie's hypothesis in 1927 at Bell Labs. For this work they shared the Nobel prize in 1937 with G. P. Thomson. Their results were published in a paper entitled "The scatterring of electrons by a single crystal of nickel" in the journal {\it{Nature}} in 1927. In their paper, Davisson and Germer reported their analysis of the angular distribution of electrons scattered from nickel. They showed that the electron beam was scattered by the surface atoms on the nickel crystal at the {\it{ exact angles}} that had been predicted for the diffraction of x-rays by Bragg's formula, with a wavelength given by the de Broglie equation (1). This was the first time that Bragg's law was applied to electrons. In the same year G. P. Thomson reported his experiments, in which a beam of electrons was diffracted by a thin foil. Thomson found patterns that resembled the x-ray patterns. 

The Davisson and Germer experiment is very simple to understand.  Electrons strike a nickel crystal which is cut parallel to a set of its ${111}$ planes[ see Fig. 1]. The kinetic energy of these electrons is controlled by the accelerating voltage, $V$. Electrons are scattered in all directions at all speeds of bombardment. The intensity of the electrons scatterred off the target at various angles was analyzed. It was seen that this intensity peaked for certain critical energies at a given scatterring angle. The Bragg condition for maximum constructive interference is
\begin{equation}
2d \sin A=m\lambda, m=1,2,......
\end{equation}
where $d$ is the spacing between the planes as shown in Figure 2, $\lambda$ is the wavelength and $A$ is the angle between the incident beam and the plane from which scatterring is taking place [see Figure 2]
From Figure 2 it is clear that this can be re-written in terms of the angle $B$ as:
\begin{equation}
2d \cos \frac{B}{2}=m\lambda, m=1,2,......
\end{equation}
and 
\begin{equation}
d=a\sin\frac{B}{2},
\end{equation}
where $a$ is the lattice spacing in the nickel crystal. This gives us
\begin{equation}
\lambda=\frac{a \sin B}{m}.
\end{equation}
For nickel, $a=0.215nm$. A peak in the electron intensity at an agle $\phi=50^{0}$ for $m=1$ gives the electron wavelength as $0.165nm$. Davisson and Germer found that at this angle the peak corresponds to a voltage $V=54 Volts$. Corresponding to this voltage, the momentum of the electron is given by
\begin{equation}
p=\sqrt{2m_{e}eV},
\end{equation}
where $m_{e}$ is the mass of the electron and $e$ is its charge. The de Broglie wavelength corresponding to this momentum is
\begin{equation}
\lambda=\frac{h}{p}=0.167nm.
\end{equation}
This was undoubtedly in excellent agreement with the experimental results. Shortly after this experiment, G P Thomson demonstrated a similar interference phenomenon with electrons. \{These two experiments were stunning validations of the de Broglie hypothesis and the understanding of the physical world took a whole new meaning - particles can also propagate like waves.}
\section{The Hitachi group's double-slit interference experiment for electrons}
While the Davisson and Germar experiment left no doubt about the wave nature of electrons, the most appealing and satisfying testimonial of the electron's wave like properties would definitely be the classic paradigm of quantum mechanics - the double slit interference experiment. As already mentioned in the introduction, the first attempts to do this go back to the late 1950s when Gottfried M\"{o}llenstedt and Heinrich D\"{u}ker of the University of T\"{u}bingen in Germany used an electron biprism to split an electron beam into two components and observe interference between them. Following this,  Claus J\"{o}nsson of the University of T\"{u}bingen did the experiment. In 1974 researchers led by Pier Giorgio Merli did the electron interference experiment at the University of Milan. The experiment was repeated in 1989 by Tonomura et al at Hitachi in Japan. By 1989 stunning advances in technology, particularly in electronics made the Hitachi group's equipment far more sophisticated, precise and elegant. In a paper entitled '{\it{ Demonstration of Single-Electron Buildup of an Interference Pattern}}' published in the {\it{ American Journal of Physics}} in 1989, Akira Tonomura and colleagues at the  Hitachi Advanced Research Laboratory in Japan reported the double slit interference experiment with {\it{ single electrons}}. In their experiment they used an electron microscope equipped with an electron biprism and a position sensitive electron counting system. In the following, we describe this experiment briefly.

 Electrons are emitted one by one from the source in the electron microscope and they encounter the biprism [see Fig. 3]. These electrons were accelerated to 50,000 Volts. Electrons having passed through on both sides of the filament were then detected one by one as particles at the detector. The detectors used were so good that even  a single electron would be detected with a hundred percent efficiency. At the beginning of the experiment bright spots began to appear - these were signatures of electrons detected one by one as particles. These bright spots in the beginning appear to be randomly positioned on the detector screen. It may be noted that {\it{ only one electron is emitted at a time}}. When a large number of electrons is accumulated over time, a pattern that looks like regular fringes begins to appear on the detector screen. After about twenty minutes very clear interference fringes could be seen -  these fringes are made up of accumulated bright spots, each of which records the detection of an electron! Each time a bright spot is seen, we understand it as an electron detected as a 'particle' and yet, the build up over time of an unmistakable inteference pattern is undoubtedly a signature of waves! Keeping in mind that that there was only one electron entering the set up at a time, the Hitachi group's experiment clearly demonstrated electrons behaving like waves as described by quantum mechanics. The interference pattern is a consequence of the possibilities of two different paths (amplitudes) for the single electron to pass through as it encounters the biprism - a situation exactly equivalent to a single electron encountering a double slit. The out of the way chance of  that the pattren is due to two electrons being together (electron-electron interaction) is completely ruled out in the experiment as the second electron is not even produced from the cathode of the electron microscope till long after the first electron is detected.

It is easy to see how the experiment implements the double-slit situation. At the heart of the Hitachi group's experiment was the {\it{ electron biprism}}. The electron biprism was invented in 1953 by Gottfried M\"{o}llenstedt. For the past five decades it has proven to be an important tool in the study of electron waves and applications in solid state physics and holds tremendous potential for applications in modern nanotechnology. Together with his Ph.D student, Heinrich D\"{u}ker, M\"{o}llenstedt developed the electron biprism. This initially consisted of a $1\mu m$ thin wire which was chargeable through a voltage source. The biprism of the kind that was used by the Hitachi group consists of two grounded plates with a fine filament between them. The filament has a positive potential with respect to the plates. The  filament used by the group was thinner than 1 micron in diameter. If the incoming electron wave is given by 
\begin{equation}
\psi = e^{ik_{z}z},
\end {equation}
the action of the  biprism is to to deflect the beam. If the electrostatic potential in the $ x z$ pane is $V(x, z)$, then the defelcted wave is:
\begin{equation}
\psi(x,z) = \exp{\Big(i k_{z}z - \frac{me}{\hbar^{2}k_{z}} \int_{-\infty}^{z} V(x, z')dz'\Big)}.
\end{equation}
In the experiment, the kinetic energy of the electrons, $\frac{\hbar ^{2} k_{z}^{2}}{2m} >> e|V(x,z)|$. There are two possible ways this wave can be defelted by the biprism, depending on which side it passes by - In each case, the deflected wave can be approximated as $e^{ik_{z}z \pm e^{ik_{x}x}}$ upto a constant factor, where 
\begin{equation}
k_{x}=- \frac{me}{\hbar^{2}k_{z}}\int_{-\infty}^{\infty} \Big( \frac{\partial V(x, z')}{\partial x} \Big)_{x=a} dz',
\end{equation}
taking into account the fact that $V(x,z)=V(-x,z)$, i.e., the potential is symmetrical. After defelction, the waves would propagate towards the centre as $k_{x}>0$. This deflection can be viewed as some sort of an impulse that each wave would experience - having the same amplitude but different signs depending on which side of the filament they pass. The overlapping of these two amplitudes in the observatiuon plane would then give rise to the wave:
\begin{equation}
\psi(x,z) = e^{k_{z}z} (e^{-ik_{x}x} + e^{ik_{x}x}).
\end{equation} 
The probability distribution corresponding to this would contain an interference term, $4 \cos^{2}(k_{x}x)$, and this is what is observed. In the Hitachi group experiment  parameters were chosen to give a fringe spacing of the pattern of the order of $900 A^{0}$. The electrons were detected using a two dimensional position sensitive electron counting system. This system comprised of a flourescent film and a photon counting image acquisition system. (For more details  on this stunning experiment, the interested reader is referred to the literature listed at the end of the article.) Some readers might be aware that in September 2002, the double-slit experiment of Claus J\"{o}nsson was voted "the most beautiful experiment" by readers of Physics World. To quote Robert Crease in an article dicussing this poll, "The double-slit experiment with electrons possesses all of the aspects of beauty......It is transformative, being able to convince even the most die-hard sceptics of the truth of quantum mechanics". Interestingly, unlike Young's double-slit experiment for light, the double-slit interference experiment for electrons has nobody's name attached to it.

The experiment by Tonomura and colleagues at Hitachi unambiguosly demonstrated the single electron interference phenomenon in all its glory, brilliantly capturing the image of the interference patterns in the now famous picture (Fig.3). 

\section{Interference experiments with atoms, molecules, bucky balls and more}

Clearly, the wave nature of matter has been demonstrated beyond doubt with the experiments mentioned and discussed in the previous sections. It is often argued that this uniquely quantum mechanical feature escapes our everyday perception because of the 'smallness' of Planck's constant, $h$ being as small as $6.6$ x $10^{-34} Js$. For a macroscopic object this would make the de Broglie wavelength so small that its quantum nature (wave-particle duality) would not be observable. However, this has been no deterrent for a large number of brave experimentalists who have verified the wave nature of matter not only for electrons but also for atoms, dimers, neutrons, molecules, nobel gas clusters and even Bose-Einstein condensates. Quantum leaps in technology and sophisticated instrumentation have made dreams of these gedanken experiments a reality. An interesting question that arises is, how far can we go with larger objects? What is the limit for observing this quantum feature in terms of size, mass, complexity? In the following we describe a recent set of experiments which demonstrate quantum interference in some of the most massive molecules by far -  $C_{60}$ and $C_{70}$ fullerenes and {\it{tetraphenylporphyrin molecules}} which are biological molecules present in chlorophyll and haemoglobin and are twice the size of fullerenes. These experiments could hold the key to answering fundamental questions  about quantum mechanics and the nature of the quantum-classical transition and more. 

$C_{60}$, the third allotropic form of carbon was discovered in 1985 by Kroto and colleagues. These carbon molecules have a structure of a truncated icosahedron.[see Fig.5.] The truncated icosahedron has 12 pentagon and 20 hexagon rings and has 60 vertices - the shape of the soccer ball. These molecules have been  called 'buckminsterfullerene' or just 'fullerenes' because of their striking resemblance to geodesic structures first discussed by Leanardo da Vinci and then implemented in architecture  by the architect Buckminster Fuller. In a paper published in {\it{ Nature}} in 1999, the group led by Anton Zeilinger in Vienna  observed de Broglie wave interference of the buckminsterfullerene $C_{60}$ - the most stable fullerene with a mass of 720 atomic units, composed of 60 tightly bound carbon atoms. This is a fascinating result - intuitively one would expect a 60 atom molecule like the fullerene to behave more like a {\it{ classical particle} than like a quantum mechanical particle!} In the following we briefly describe this experiment.

Fullerene molecules were brought into the gas phase by sublimating the powder form in an oven at  a temperature of approximately 900K. Molecules are ejected one by one through a small slit in the oven. The de Broglie wavelength of these molecules (uniquely determined by the momentum of the molecule) is $\lambda=2.8pm$. It turns out that the de Broglie wave length is approximately 400 times smaller than the size of the particle! The interference pattern expected would therefore be very small and very sophisticated machinery will be needed to see it. The diffracting element used by the group was a nanofabricated free standing silicon nitride grating with a grating constant $d=100nm$ and a slit openeing of approximately $50nm$. After free evolution over 1 meter,  the fullerene molecules are detected via thermionic ionization by a tightly focused Argon ion laser beam operating at $24 W$. The positive ions are counted by a secondary electron counting system. The counts, as a function of position clearly showed a diffraction pattern. Note that just as in the case of the Hitachi experiment, the pattern is built up atom by atom. The experiment  ensures that there is no interference between two or more particles during their evolution in the apparatus - so this is indeed a single particle quantum phenomenon.

From $C_{60}$, the group has gone on to repeat the experiment for larger, more complex molecules. Starting first with the $C_{70}$ fullerene, the group demonstrated this remarkable phenomenon in $C_{60}F_{48}$ - a fluorofullerene - which at $1632$ atomic units is the largest and most complex molecule till date to show quantum interference. The group also demonstarated quantum interference for tetraphenylporphyrin, a derivative of a biodye which is found in chlorophyll. This is the first biomolecule exhibiting wave nature and has a spatial extent of $2nm$ - almost twice as much as $C_{60}$. There is hardly any need to emphasize that these large molecules are, in many respects, like classical objects. They can store a lot of internal energy in many degrees of freedom. When heated to about 3000 K, fullerenes can emit electrons, photons and even diatomic carbon molecules. This is similar to a hot object glowing and emitting black-body radiation. This makes these experiments even more remarkable as they achieved nothing less than capturing the underlying quantum footprints - the wave-particle duality - of large and complex classical-like objects. This begs the question - is there a limit to the size and complexity of the object that can show quantum interference? This question leads us to the age old debate about the classical- quantum transition and the connection between these two complelety different descriptions of reality. It is often argued that quantum mechanics is the description for an abstract {\it{ micro}} world far removed from reality while classical mechanics describes the physics of the {\it{ macro}}world of our everyday experience. But the {\it{ macro}} is finally composed of the {\it{ micro}}! Where then, is the boundary, if any? These spectacular experiments offer the tantalizing possibility of probing the twilight zone between quantum and classical worlds by performing interference experiments with increasingly heavier and complex objects.

\subsection{The quantum-classical boundary and decoherence}
 A widely accepted explanation for the appearance of classical like features from an underlying quantum world is the {\it{ environment induced decoherence}} approach. According to this theory, coupling to a large number of degrees of freedom (the environment) results in a loss of quantum coherence which leads to emergent classicality. In the context of the experiments described above for fullerenes and other large molecules, an important decoherence mechanism comes from its interaction with particles from the background gas. By flooding the interferometer with various gases at low pressure Anton Zeilinger's group studied the effect of decoherence on the inteference phenomenon. In fact, in keeping with the theoretical predictions, they saw an exponential decrease of the observed fringe visibility. It is interesting to note that such decoherece which is caused by collisions is almost impossible to test in the usual matter wave interferometry with smaller particles (like electrons and neutrons) as the particles are themselves so light that they would be kicked out of the interferometer after colliding with a gas particle. In the case of fullerenes and larger molecules, the molecules themselves are  heavy enough to remain in the interferometer after a typical collision. Apart from confirming the qualitative and quantitative predictions of the decoherence theory, these  experiments allow one to estimate the vacuum conditions that are required for the successful observation of quantum interference of much larger objects. The surprising observation by the group was that collisions would not limit quantum interference even for an object as large as a virus provided the background pressure of the gas is reduced to below $3$ x $10^{-10}$ mbar.

\section{Conclusions}
The matter-wave interference experiments of massive molecules described above have allowed us to probe and explore a new regime of Nature and opened up the possiblity of experimentally studying the elusive quantum-classical boundary. These stunning studies have demonstrated beyond doubt that the quantum nature of large objects can indeed be captured experimentally in the classic paradigm of the double slit interference and diffraction set ups. Important decoherence mechanisms have been studied and identified and the good news is that it is possible to carry these experiments further for heavier and more complex molecules. Infact, there is talk of doing these interference experiemnts for proteins like Insulin and then on to larger proteins, clusters and nanocrystals. In the last two decades matter wave interferrometry have demonstrated effects that were previously unthinkable. More importantly, they have opened up exciting possibilities of exploring questions of fundamental interest in the foundations of quantum mechanics - like the quantum-classical boundary. Imaginative and novel ideas continue to fuel the field and it can be safely said that in our pursuit of the "only mystery of quantum mechanics" the best and the most interesting experiments are yet to come.

\begin{flushleft}
{\bf Suggested Reading}
\end{flushleft}
\noindent {\bf The topics touched upon in this article cover several  references. The interested reader may look at some of the following:}

\begin{enumerate}

\item{{\it {The Feynman Lectures on Physics (Vol-III)}}, R. P. Feynman, Robert B Leighton and Mathew Sands, Addison Wesley (1964)}

\item{{\it {The Double-Slit Experiment}}, Physics World, September 2002, p15.
An extended version of this article including three letters about the history of the double-slit experiment with single electrons is available at\\ {\bf http://physicsworld.com/cws/article/print/9745}}

\item{{\it {Demonstration of single electron build up of an interference pattern}}, A. Tonomura et al, American Journal of Physics, {\bf 57}, 2, February 1989. A nice description of this experiment can also be found at the Hitachi web site: \\
http://www.hqrd.hitachi.co.jp/global/doubleslit.cfm}

\item{A non technical description of the Fullerene diffraction experiments can be found at the web site of Prof. Anton Zeilinger's Research group at the  Universit\"{a}it Wien, Austria: \\
http://www.quantum.univie.ac.at/research/matterwave/c60/index.html}

\item{{\it {The Coming of a Classical World}}, Anu Venugopalan, {\bf Resonance: Journal of Science Education,Vol 9, 10,September 2004}}
\end{enumerate}

\pagebreak
\begin{figure}
\begin{center}
\epsfig{figure=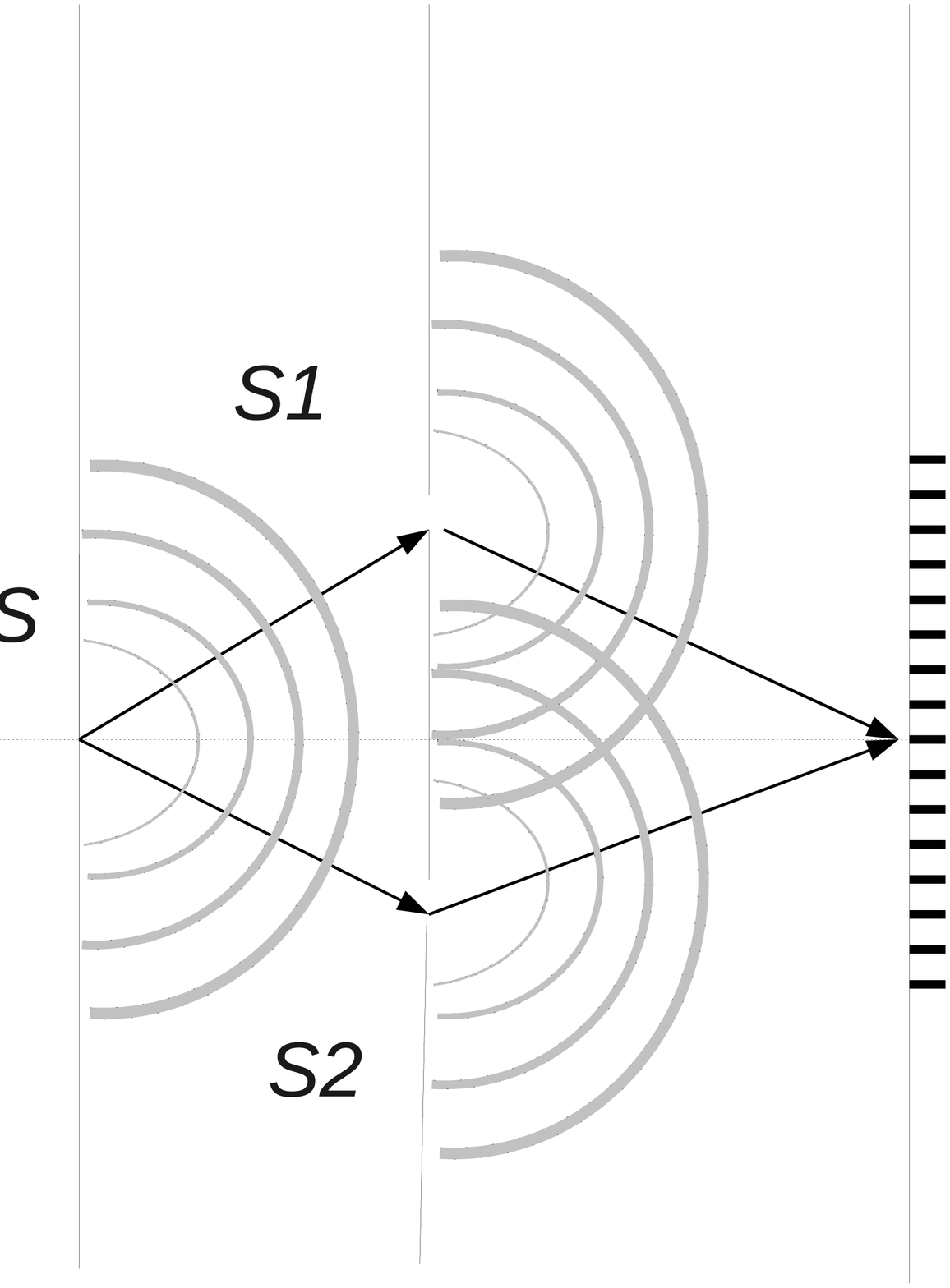}
\vspace{0.5 in}
\caption{\bf The double slit interference experiment}
\end{center}
\end{figure}
\begin{figure}
\begin{center}
\epsfig{figure=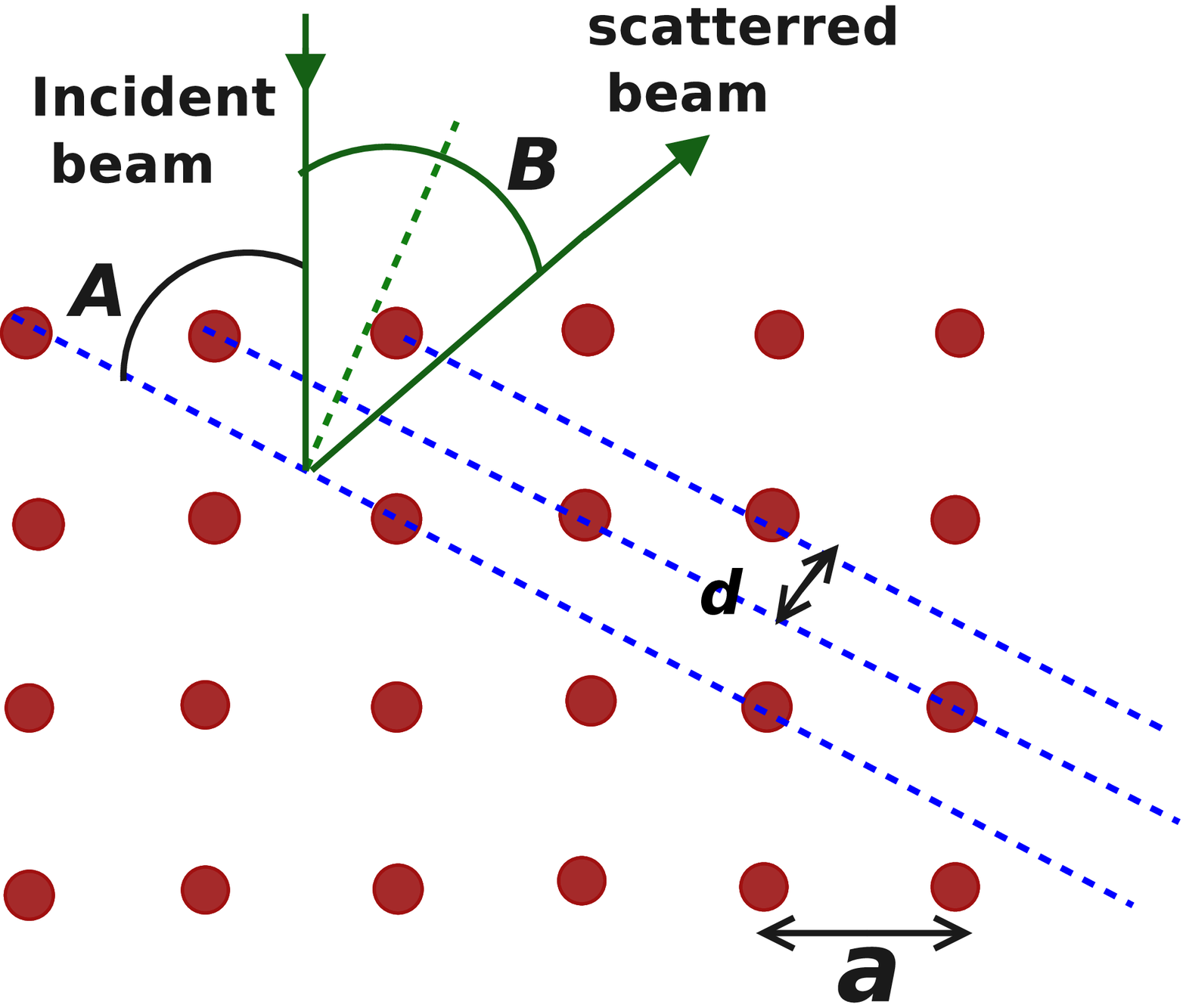}
\vspace{0.5 in}
\caption{\bf The Davisson and Germer experiment}
\end{center}
\end{figure}
\begin{figure}
\begin{center}
\epsfig{figure=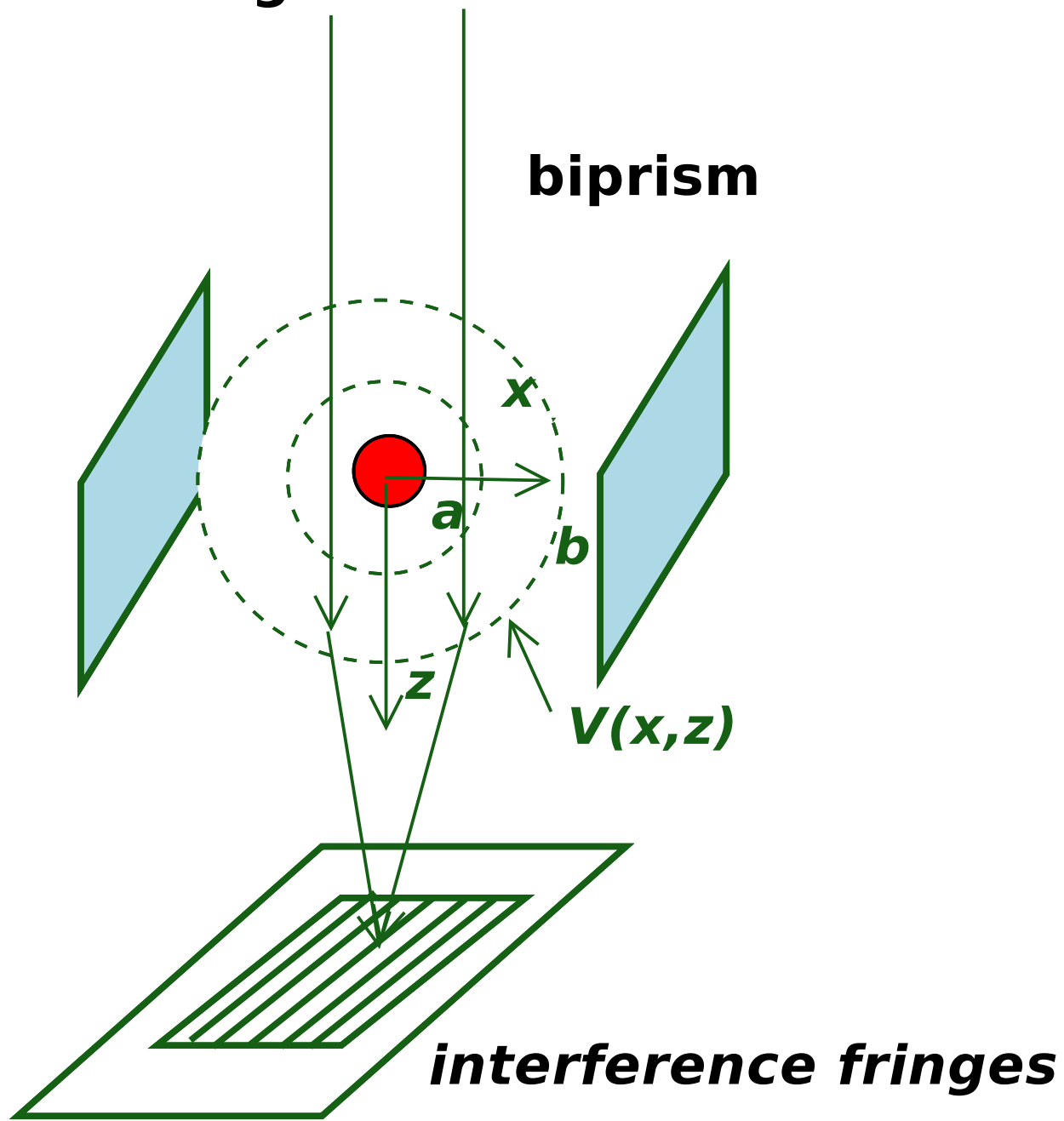}
\vspace{0.5 in}
\caption{\bf Set up for double-slit interference with single electrons}
\end{center}
\end{figure}
\begin{figure}
\begin{center}
\epsfig{figure=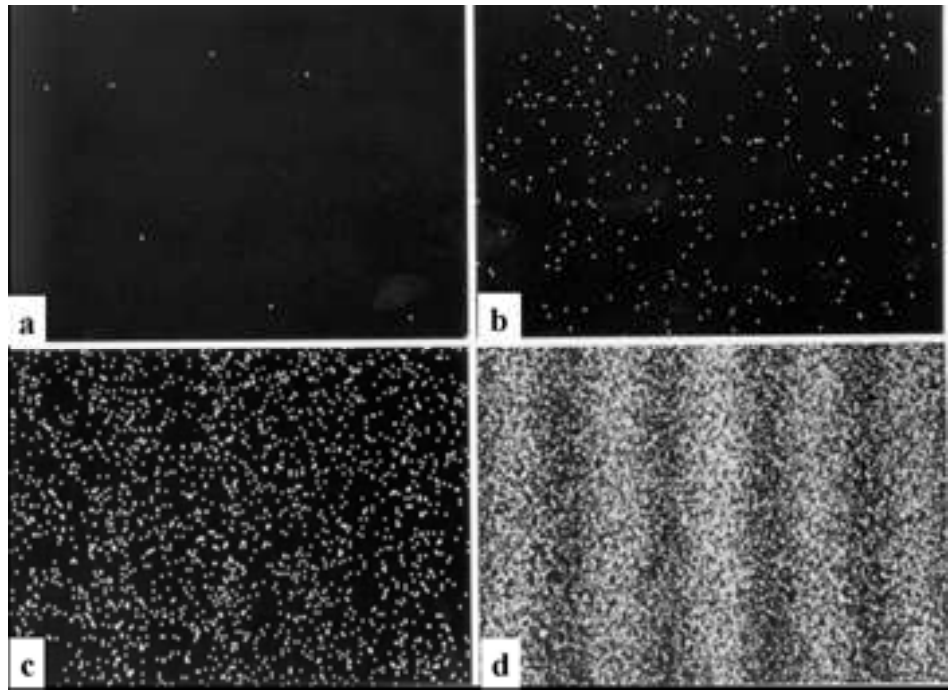}
\vspace{0.5 in}
\caption{\bf Single electron events build up to from an interference pattern in the double-slit experiments:The number of electron accumulated on the screen. (a) 8 electrons; (b) 270 electrons; (c) 2000 electrons; (d) 160,000. The total exposure time from the beginning to the stage (d) is 20 min. (Reproduced from http://www.hqrd.hitachi.co.jp/em/doubleslit.cfm, with permission from the authors)}
\end{center}
\end{figure}
\begin{figure}
\begin{center}
\epsfig{figure=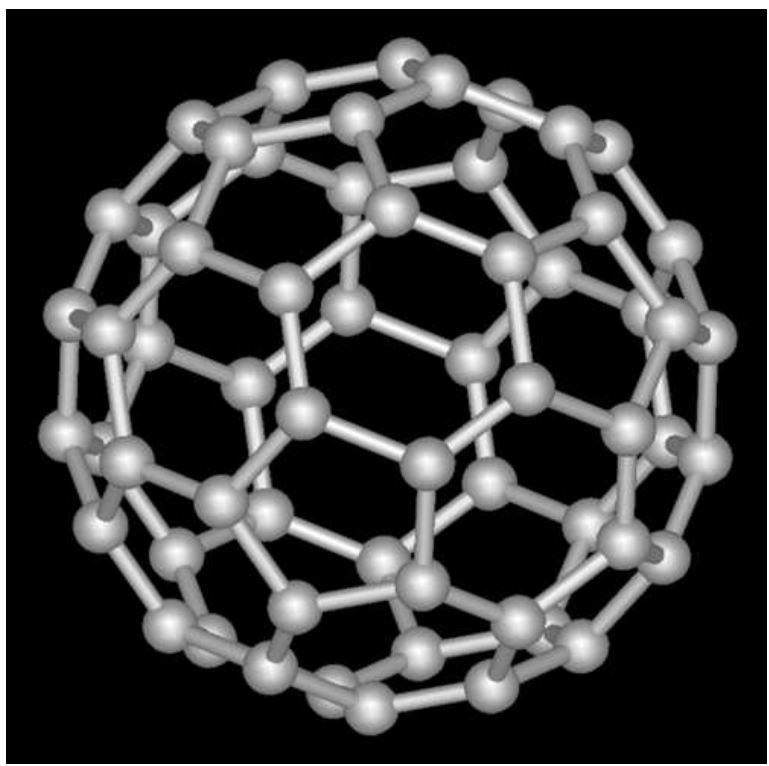}
\vspace{0.5 in}
\caption{\bf The C-60 Fullerene molecule (picture dowloaded from http://commons.wikimedia.org/wiki/Image:Fullerene-C60.png}
\end{center}
\end{figure}
\end{document}